\documentclass[prb,twocolumn,amsmath,showpacs,preprintnumbers,superscriptaddress,amssymb,space,data]{revtex4}
\usepackage{amssymb}
\usepackage{makeidx}
\usepackage{amsmath}
\usepackage{dcolumn}
\usepackage{bm}
\usepackage{graphicx}
\usepackage{amsfonts}

\begin{document}

\title{Amplitude and phase effects in Josephson qubits driven by a biharmonic electromagnetic field}

\author{A. M. Satanin}
\email{sarkady@mail.ru}

\author{M. V. Denisenko}

\affiliation{Nizhny Novgorod State University, 23 Gagarin Ave., 603950, Nizhny Novgorod, Russia }

\author{A. I. Gelman}

\affiliation{Institute of Applied Physics RAS, 46 Ul'yanov Str., 603950, Nizhny Novgorod, Russia}

\author{Franco Nori}

\affiliation{CEMS, RIKEN, Saitama, 351-0198, Japan}
\affiliation{Department of Physics, University of Michigan, Ann Arbor, MI 48109-1040, USA}

\begin{abstract}

We investigate the amplitude and phase effects of qubit dynamics and excited-state population under the influence of a biharmonic control field. It is demonstrated that the biharmonic driving field can have a significant effect on the behavior of quasi-energy level crossing as well as on  multi-photon transitions. Also, the interference pattern for the populations of qubit excited states is sensitive to the signal parameters. We discuss the possibility of using these effects for manipulating qubit states and calibrating nanosecond pulses.

\pacs{03.65.-a, 03.67.Hk}
\keywords{qubits; dynamic control; quasi-energy; amplitude spectroscopy}

\end{abstract}

\maketitle

\section{Introduction}
Numerous works have recently been devoted to theoretical and experimental investigations of Josephson qubit circuits (see, e.g., reviews Ref.~\onlinecite{Clarke,You,You1,Buluta}).
Amplitude spectroscopy \cite{Oliver,Berns,Sillanpaa,Wilson,Izmalkov,Berns1,Rudner} obtains information about these circuits as a function of the driving amplitude and control parameters determining the distance between levels. This technique can be applied to quantum systems with crossing energy levels where transitions can be realized by changing the external parameters. In this situation, the frequency of the applied electromagnetic field can be several orders of magnitude lower than the distance between levels, thus the system driven at the field period evolves mostly adiabatically, with the exception of the relatively small time intervals when energy levels approach each other and Landau-Zener tunneling becomes possible between them \cite{Landau,Zener,Majorana}. This makes it possible to obtain an interference pattern of populations depending on the field amplitude and the distance between levels \cite{Stuckelberg} (see Ref.~\onlinecite{Shevchenko} for a review). The main advantage of amplitude spectroscopy is that the system can be investigated in a wide range of field changes and inter-level distances (level displacements) and also provides information about the effects of noise on a qubit.

Many problems of qubit dynamics are not fully solved at present. These include the problem of reducing the effect of different noise mechanisms \cite{You,You1}, optimal control \cite{Sporl}, and nonlinear dynamics of qubits \cite{Oliver,Berns,Sillanpaa,Wilson,Izmalkov,Berns1,Rudner}, etc. It has been known that  high-frequency pulses with Rabi frequency can be used to control the dynamic of qubits. Meanwhile, the dynamics of a qubit is not determined by the field produced by a pulse generator but by the acting field, which undergoes significant changes in a waveguide. In recent works Ref.~\onlinecite{Bylander,Gustavsson} the control of qubit populations and signal diagnostics were carried out by mixing two large-amplitude RF-pulses with different frequencies at a fixed the phase difference.
In particular, the ability to manipulate pulse shapes can be used to control the time a qubit spends near an avoid crossing. This approach in combination with Landau-Zener-St\"{u}eckelberg interference, can control the interference, by changing the parameters of a probing signal \cite{Sun1,Wang,Sun,Sun3,Shevchenko1,Shevchenko2}.

Since Rabi dynamics and qubit populations depend on the form of the driving signal, the qubit could also be used for calibrating ultrashort (nanosecond) pulses. For example, phase-sensitive effects, actively used in optics \cite{Paspalakis,Nerush} and plasma physics \cite{Silaev}, can control system populations and calibrate ultrashort laser pulses. Biharmonic drives have also been extensively studied in the context of controlling transport phenomena of either small particles or magnetic flux quanta \cite{Savel'ev}.

The main goal of this work is to investigate how to control transitions between qubit states and the interference pattern of populations by changing the form of the applied driving field. A perturbation resonant theory \cite{Scully} (Rabi generalized approximation) and a quasi-energy approach \cite{Shirley,Quasienergy,Grifoni,Son} are used to study the controlled dynamics of qubits subject to driving. Special attention is paid here to the phase dependence of the qubit response to a biharmonic field, which represents the superposition of two signals with a phase shift between them. We describe interesting phase effects, which can be observed in Josephson circuits by means of amplitude spectroscopy, when qubits are driven by biharmonic signals.

This work is organized as follows. At first, we describe a model of a Josephson loop using a two-level approximation, explain the meaning of the control parameters and analyze the qubit dynamics driven by biharmonic pulses by using the rotating-wave approximation (RWA). Then the Floquet formalism and approach \cite{Grifoni} based on the quasi-energy representation for transition probabilities is briefly described. Further we present the results of numerical calculations and their analysis based on the RWA. Finally, we discuss several consequences of our analysis.

\section{BIHARMONICALLY-DRIVEN QUBIT MODEL}
\begin{figure}[t]
\begin{center}
      \includegraphics[width=6.5cm,height=6.5cm]{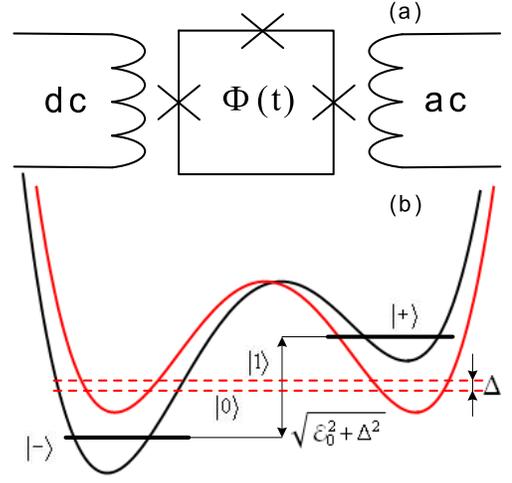}
\end{center}
\caption{\label{fig1} (color online) Schematic diagram (a) of a flux qubit in driving field and (b) the qubit levels in the effective potential. The red curve depicts a potential profile with no displacement $\varepsilon(t)=0$, while the black curve gives the levels at the static bias $\varepsilon(t)=\varepsilon_{0}$.}
\end{figure}

The basic dynamical behavior of a superconducting flux qubit driven by an electromagnetic field can be described by the Hamiltonian
%
%
\begin{equation}
H(t)=\frac{1}{2}\begin{pmatrix}
                 \varepsilon(t) & \Delta \\
                 \Delta & -\varepsilon(t)
                 \end{pmatrix}, \label{1}%
\end{equation}
where $\varepsilon(t)$ is the energy bias of the qubit, and $\Delta$ is the tunnel level splitting  \cite{You, You1}. The qubit may be driven with an external magnetic flux $\Phi(t)$ consisting of constant and alternating fluxes $\Phi(t)=\Phi_{dc}+\Phi_{ac}(t)$ [see Fig.~\ref{fig1}(a)]. In this case, the energy bias $\varepsilon(t)=\varepsilon_{0}+\varepsilon_{\sim}(t)$ describes the time-dependent driving
\begin{equation}
\varepsilon_{\sim}(t)=2I_{p}\Phi_{ac}(t),\label{1.1}
\end{equation}
with the static bias
\begin{equation}
\varepsilon_{0}=2I_{p}(\Phi_{dc}-\Phi_{0}/2),
\end{equation}
where $I_{p}$ is persistent current, $\Phi_{0} = h/2e $ is the magnetic flux quantum. We shall treat below $\varepsilon_{0}$ as a controlling parameter. When only the dc-magnetic flux, $\Phi_{\textrm{dc}} = \Phi_{0}/2$, penetrates the superconducting circuit, then the potential energy of the qubit becomes a double-well potential \cite{Orlando} [depicted in Fig.~\ref{fig1}(b) by a red curve]. In this static case, quantum mechanical tunneling causes the appearance of two discrete levels, the qubit, with energies $E_{0}=-\Delta/2$ and $E_{1}=\Delta/2$, characterized by the corresponding basis vectors $|0\rangle=\frac{1}{\sqrt{2}}\left(\begin{matrix}
                                   1\\-1
                                   \end{matrix}\right)$
and $|1\rangle=\frac{1}{\sqrt{2}}\left(\begin{matrix}
                                   1\\1
                                   \end{matrix}\right)$.
The states $|0\rangle$ and $|1\rangle$ are coherent superpositions of states with electrical currents flowing clockwise and counter-clockwise in the superconducting circuit. Changing the external magnetic flux $\Phi_{\textrm{dc}}$ modifies the effective potential and states, $|\pm\rangle$, of the qubit with energies $E_{\pm}=\pm\frac{1}{2}\sqrt{\varepsilon_{0}^{2}+\Delta^{2}}$.

To perform quantum control we consider the driving function to be periodic in time $\varepsilon(t)=\varepsilon(t+T)$. Although our approach is applicable to any periodic function $\varepsilon(t)$, here we shall discuss in detail the case of a \textit{biharmonic drive}
\begin{equation}
\varepsilon_{\sim}(t)=A\left[\cos(\omega t)+\gamma\cos(2\omega t+\theta)\right], \label{2}%
\end{equation}
where $A$ is the driving amplitude parameterized in units of energy, $\theta$ is the relative phase of the signals, and $\gamma$ is the relative amplitude.
Note, that in an experiment \cite{Bylander} the signal generator allows one to control the relative signal phase.

The system dynamics obeys
%
%
\begin{equation}
i\hbar\frac{\partial}{\partial t} |\psi(t)\rangle=H(t)|\psi(t)\rangle. \label{3}%
\end{equation}
Using the RWA \cite{Scully}, let us now investigate the system behavior described by the Hamiltonian (\ref{1}), where $\varepsilon(t)$ is given by Eq.~(\ref{2}). We perform the canonical transformation
%
%
%
\begin{equation}
|\psi(t)\rangle=U_{0}(t)|\overline{\psi}(t)\rangle,\quad U_{0}(t)=\!\exp\!\!\left[\!-\frac{i}{2\hbar} \phi(t)\sigma_{z}\right], \label{4}%
\end{equation}
where
%
%
\begin{center}
\begin{equation}
\begin{split}
\phi(t)=\varepsilon_{0}t+\frac{A}{\hbar\omega}&\left[\sin(\omega t)+\frac{\gamma}{2}\left(\sin(2\omega t +\theta)-\sin\theta\right )\right ],{}\\
&\sigma_{z}=\left(\begin{matrix}
           1 & 0 \\
           0 & -1
           \end{matrix}\right). \label{5}%
\end{split}
\end{equation}
\end{center}
The Schr\"{o}dinger equation for the transformed wave function $|\overline{\psi}(t)\rangle$ takes the form
%
%
\begin{equation}
i\hbar\frac{\partial}{\partial t}|\overline{\psi}\rangle=\left(U^{+}_{0}H(t)U_{0}-iU_{0}^{+}\frac{\partial U_{0}}{\partial t}\right )|\overline{\psi}\rangle=\overline{H}(t)|\overline{\psi}\rangle, \label{6}
\end{equation}
and the modified Hamiltonian becomes
%
\begin{equation}
\overline{H}(t)=\frac{\Delta}{2}\sum^{\infty}_{n, m =-\infty}J_{n}\!\left(\frac{A}{\hbar\omega}\right)
\!J_{m}\!\left(\frac{\gamma A}{2\hbar\omega}\right)
\left (
\begin{matrix}
0 & d_{+}^{(n,m)}(t)\\
d_{-}^{(n,m)}(t) & 0
\end{matrix}
\right ),\label{7}%
\end{equation}
where

$ d_{\pm}^{(n,m)}(t)\!=\!\exp\!\left(\mp i\left[\frac{A\gamma}{2\hbar\omega}\sin\theta\!+\!m\theta\!\right]\!\right)\!\exp\!\left(\pm i[\frac{\varepsilon_{0}}{\hbar}\!+\!(n\!+\!2m)\omega]t\right)$.
To obtain Eq.~(\ref{7}) a well-known relation was used
%
%
\begin{equation}
\exp\!\!\left(i\frac{A}{\hbar\omega}\sin(\omega t)\right)=\sum_{n}J_{n}\!\!\left(\frac{A}{\hbar\omega}\right )\!\exp(in\omega), \label{8}%
\end{equation}
where $J_{n}(x)$ is a Bessel function. Using the RWA in Eq.~(\ref{7}), fast-oscillating components can be neglected with the exception of those for which the resonance condition is held:
$\varepsilon_{0}+(n+2m)\hbar\omega=0$, at $\hbar\omega\gg\Delta$. Then the Hamiltonian describing the slow dynamics will have the form
%
%
\begin{equation}
\overline{H}_{R}=\frac{1}{2}\begin{pmatrix}
                                             0 & \Delta_{R}\\
                                             \Delta^{*}_{R} & 0
                                             \end{pmatrix}, \label{9}
\end{equation}
where the resonance parameter is introduced
%
%
\begin{multline}
\Delta_{R}\equiv\Delta_{R}(A,\gamma,\theta)=\Delta\exp\!\!\left(-i\frac{A\gamma}{2\hbar\omega}\sin \theta\right)\cdot\\
\sum_{n,m}J_{n}\!\!\left(\frac{A}{\hbar\omega}\right)J_{m}\!\!\left(\frac{A\gamma}{2\hbar\omega}\right)\exp(-im\theta) \label{10}
\end{multline}
and the sum is taken over all $n$ and $m$ satisfying the condition $\varepsilon_{0}+(n+2m)\hbar\omega=0$.  If the amplitude ratio $\gamma$ or the phase $\theta$ is fixed and the definite value of the control parameter  $\varepsilon_{0}$ is also chosen, then [according to the resonance condition $\varepsilon_{0}+(n+2m)\hbar\omega=0$]  it is possible to find a set of values of $n$ and $m$ for the Bessel function products in the expression (\ref{10}), which determine the character of the Rabi frequency dependence on the relative amplitude $\gamma$ or phase $\theta$.

The Hamiltonian (\ref{9}) corresponds to the resonant interaction of the alternating field with a two-level system and describes a generalized Rabi resonance (see Ref.~\onlinecite{Scully}). When $\gamma=0$ this expression reduces to the standard Rabi resonance ($m=0$) in the case of a monochromatic signal \cite{Ashhab}. For a biharmonic signal, the frequency of the generalized Rabi resonance is defined by the expression $\Omega_{R}=\left|\Delta_{R}\right|$, which depends on the amplitude driving $A$, the relative amplitude $\gamma$ and phase $\theta$ difference of the biharmonic field.

\section{QUASI-ENERGY STATES}
Let us assume that the qubit was originally in the state $|\alpha\rangle=|-\rangle$, which is the eigenvector of the Hamiltonian (\ref{1}) in the absence of  the oscillating components of the field [$\Phi_{ac}(t)=0$ in the expression (\ref{1.1})], i.e. the qubit was ``prepared'' in the ground state $E_{-}=-\frac{1}{2}\sqrt{\varepsilon^{2}_{0}+\Delta^{2}}$ [see Fig.~\ref{fig1}(b)]. We will be interested in the probability of the qubit transition to the final state $|\beta\rangle=\left ( \begin{matrix}
                                                                   1\\0
                                                                   \end{matrix}
                                                                   \right )$
(after the effect of the biharmonic drive), which is connected with the experimentally-measured current projection in the superconducting loop.
Note that this transition has been studied experimentally in Ref.~\onlinecite{Bylander,Gustavsson}.

We use the quasi-energy representation \cite{Shirley,Quasienergy} (see Ref.~\onlinecite{Grifoni} for a review) to calculate the population probabilities of the system levels. This representation provides precise intermediate states of the driven system with an optional amplitude and allows to reveal resonance transition features caused by the quasi-energy levels motion and crossing.

A formal solution of the Schr\"{o}dinger Eq.~(\ref{3}) can be written as $|\psi(t)\rangle = U(t,t_{0})|\psi(t_{0})\rangle$, where
%
%
\[
U(t,t_{0})=\hat{P}\exp\left(-\frac{i}{\hbar}\int_{t_{0}}^{t}\!\!\!H(\tau)d\tau\right ),
\]
and $\hat{P}$ denotes the time-ordering operator. The time evaluation for a period is given by the operator
%
%
\[
U(T)\equiv U(t+T, t)=\hat{P}\exp\left(-\frac{i}{\hbar}\int_{t}^{t+T}\!\!\!\!\!\!\!\!\!H(\tau)d\tau\right)
\]
which is called the Floquet operator \cite{Shirley,Quasienergy,Grifoni}. The eigenvalues of the Floquet operator can be written in the form
%
%
\begin{equation}
U(T)|\Phi_{k}(t)\rangle=e^{-iQ_{k}T/\hbar}|\Phi_{k}(t)\rangle, \quad |\Phi_{k}(t+T)\rangle=|\Phi_{k}(t)\rangle, \label{13}
\end{equation}
and the parameters $Q_{k}$ are called the quasi-energies (in the system considered here: $k=1,\, 2$). The eigenvalues $Q_{k}$ therefore can be mapped into the first Brillouin zone, obeying $-\hbar\omega/2 < Q_{k} < \hbar\omega/2$.

In the quasi-energy basis $|\Phi_{k}(t)\rangle$, the transition probability $P_{|\alpha\rangle\rightarrow |\beta\rangle}(t, t_{0})$ is described by
%
%
\begin{equation}
P_{|\alpha\rangle\rightarrow |\beta\rangle}(t, t_{0}) = \sum_{k,l} e^{-i(Q_{k}-Q_{l})(t-t_{0})/\hbar} M_{k}(t,t_{0})M^{*}_{l}(t, t_{0}), \label{14}
\end{equation}
where
%
%
\[
M_{k}(t,t_{0})=\langle\beta|\Phi_{k}(t)\rangle\langle\Phi_{k}(t_{0})|\alpha\rangle.
\]
It is clear from Eq.~(\ref{14}) that with the change of the duration of the signal $(t-t_{0})$, the contributions with different $k$ and $l$ oscillate strongly and this reduces the transition probability. When the   system parameters are changed (for example, the field amplitude, $A$, or control parameter $\varepsilon_{0}$) it is possible that two quasi-energies approach degeneracy, $Q_{k}=Q_{l}$, and  the transition probability significantly increases because it has a time-independent contribution.
In general, the crossing of quasi-energies plays an important role in populating the levels of complex quantum systems \cite{Satanin}.

It is necessary to average the expression (\ref{14}) over the initial times $t_{0}$ of the field pulse arrival at the qubit and over the biharmonic drive duration itself at the fixed signal phase \cite{Shirley}. It can be shown that the averaged transition probability $\overline{P}_{|\alpha\rangle\rightarrow |\beta\rangle}$ is determined by the relation:
%
%
\begin{equation}
\overline{P}_{|\alpha\rangle\rightarrow |\beta\rangle} = \sum_{k}\sum_{n,l}\left |\langle\beta|\Phi^{(n-l)}_{k}\rangle\right |^{2}\left |\langle\Phi^{(n)}_{k}|\alpha\rangle\right |^{2}, \label{15}
\end{equation}
where $|\Phi^{(n)}_{k}\rangle$ are the Fourier components of the quasi-energy function, which may be calculated as
%
%
\[
|\Phi^{(n)}_{k}\rangle = \frac{1}{T}\int_{0}^{T}\!\!\exp(in\omega t)\;|\Phi_{k}(t)\rangle dt.
\]

We numerically obtained the quasi-energy levels and the corresponding eigenfunctions. These can be used to find the transition probabilities in an arbitrary strong driving field and to investigate the  population dependence from different signal parameters.

\section{QUASI-ENERGY LEVELS AND MULTIPHOTON RESONANCES}
\begin{figure}[b]
\begin{center}
      \includegraphics[width=9cm,height=8cm]{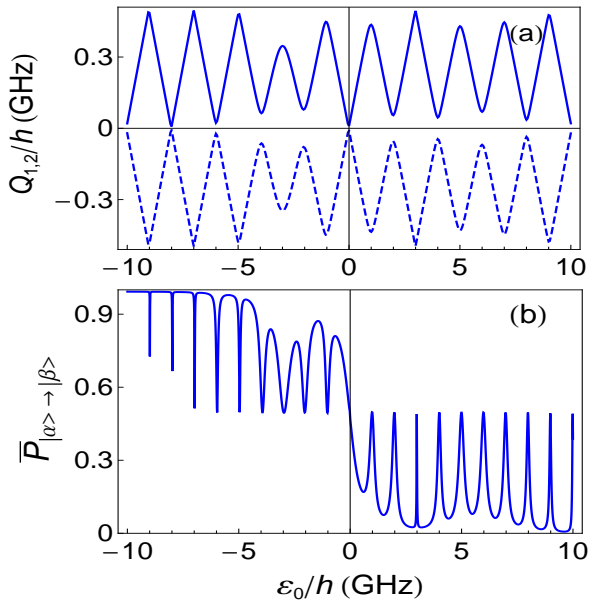}
\end{center}
\caption{\label{fig2} (color online) Quasi-energy levels ($Q_{1}, Q_{2}$) versus the displacement parameter $\varepsilon_{0}$ (a). The dashed curve corresponds to $Q_{1}(\varepsilon_{0})$ and the solid to $Q_{2}(\varepsilon_{0})$. Here the transition probability $\overline{P}_{|\alpha\rangle\rightarrow |\beta\rangle}$ versus the static bias $\varepsilon_{0}$ is shown in (b). The system parameters used here are: $\Delta/h = 0.5$ GHz, $\omega/2\pi = 1$ GHz, $A/h = 5$ GHz, $\gamma = 0.5$, and $\theta = \pi$.}
\end{figure}

We focus here on the phase dependence of the qubit excitation level population. Phase control arises by setting a relative phase difference  $\theta$ between the two components of the biharmonic drive.
First, we investigate the behavior of the quasi-energy curves $Q_{1}(\varepsilon_{0})$ and $Q_{2}(\varepsilon_{0})$, which depend on the control parameter $\varepsilon_{0}$ [see Fig.~\ref{fig2}(a)]. In the case of a biharmonic drive, the characteristic feature of the $Q_{k}(\varepsilon_{0})$ functions is symmetry-breaking: $Q_{k}(\varepsilon_{0})\neq Q_{k}(-\varepsilon_{0})$. These features immediately follow from the expressions (\ref{1}), (\ref{2}), and are clearly observed in Fig.~\ref{fig2}(a). For some particular values of the control parameter $\varepsilon_{0}$, when the resonance conditions are fulfilled, $\varepsilon_{0}+(n+2m)\hbar\omega=0$, the quasi-energy levels approach each other, causing the appearance of peaks on the diagram of the  excited level population of a qubit $|\beta\rangle$ [Fig.~\ref{fig2}(b)], which are physically specified by multiphoton transitions.
%
%
%
%

The behavior of the quasi-energies with field amplitude $A$ can also be qualitatively understood in the context of the RWA. Indeed, in this approximation, the quasi-energies are eigenvalues of the Hamiltonian (\ref{9}), i.e. they are determined by the Rabi frequency ($Q_{1}=\Omega_{R}/2$, $Q_{2}=-\Omega_{R}/2$). Thus, the expression (\ref{10}) approximately describes the dependence of the quasi-energies on the field amplitude.
\begin{figure}[t]
\begin{center}
      \includegraphics[width=9cm,height=8.5cm]{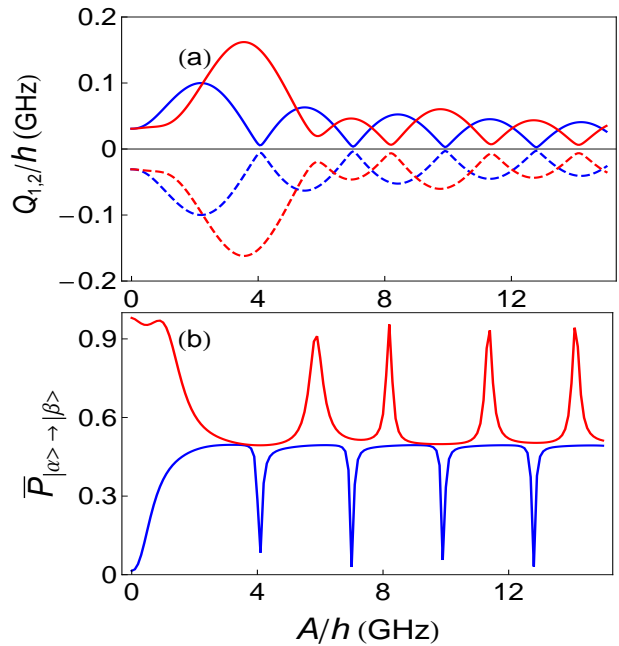}
\end{center}
\caption{\label{fig3} (color online) Quasi-energy levels ($Q_1$ and $Q_2$) in (a) and the probability, $\overline{P}_{ |\alpha\rangle\rightarrow |\beta\rangle}$, to find the qubit in the state $|\beta\rangle$ in (b). Both versus the applied-drive amplitude $A$. Here the red curves correspond to $\varepsilon_{0}/h = -2$ GHz, and the blue curves to $\varepsilon_{0}/h = 2$ GHz. The dashed curves correspond to $Q_{1}(A)$ and the solid curves to $Q_{2}(A)$. The other system parameters used here are: $\Delta/h = 0.5$ GHz, $\omega/2\pi = 1$ GHz, $\gamma = 0.5$, and $\theta=\pi$.}
\end{figure}

Figure~\ref{fig3}(a) shows the dependence of the quasi-energy levels [obtained by numerically solving Eq.~(\ref{13})]. Figure~\ref{fig3}(b) illustrates the population probabilities of the excited level [calculated according to Eq.~(\ref{15})] when changing the biharmonic field amplitude $A$. It is evident that the anti-crossing points correspond to small Rabi frequencies, which agree with the dynamic trapping (dynamic localization) of the system states. When changing the sign of the control parameter $\varepsilon_{0}$, a shift is observed where the quasi-energies approach each other due to the apparent asymmetry of the quasi-energy levels. The asymmetry of the quasi-energy levels causes the asymmetry of the excited level population as a function of the field amplitude in Fig.~\ref{fig3}(b). Note that in the case of a monochromatic driving field ($\gamma=0$) the quasi-energy levels at $\varepsilon_{0}>0$ and $\varepsilon_{0}<0$ coincide.

The analysis performed demonstrates the sensitivity of the qubit population, i.e., the measurable response of our interferometer to the form of the biharmonic field. This allows controlling the transitions between qubit levels by changing the biharmonic drive parameters.

Note that for a positive $\varepsilon_{0}$, the probability $\overline{P}_{|\alpha\rangle\rightarrow |\beta\rangle}$ of the excited level population cannot exceed 0.5 \cite{Scully}, when $\varepsilon_{0}<0$ the probability lies in the range $0.5 \leq \overline{P}_{|\alpha\rangle\rightarrow|\beta\rangle}\leq1$. Also, the curves change shape [``peaks'' are replaced by ``dips'' as seen in Figs.~\ref{fig2}(b) and ~\ref{fig3}(b)]. These results can be explained by measuring the current projection in the superconducting loop. Thus, for the opposite-current projection, i.e. $|\beta\rangle=\left ( \begin{matrix}
                              1\\0
                              \end{matrix}
                              \right )$
when $\varepsilon_{0}>0$, the probability does not exceed 0.5 (i.e., $0 \leq \overline{P}_{ |\alpha\rangle\rightarrow |\beta\rangle}\leq0.5$). This is why we describe the character of the resonances (``peaks'' and ``dips'') according to their forms for positive $\varepsilon_{0}$.

\begin{figure}[t]
\begin{center}
      \includegraphics[width=8.5cm,height=13cm]{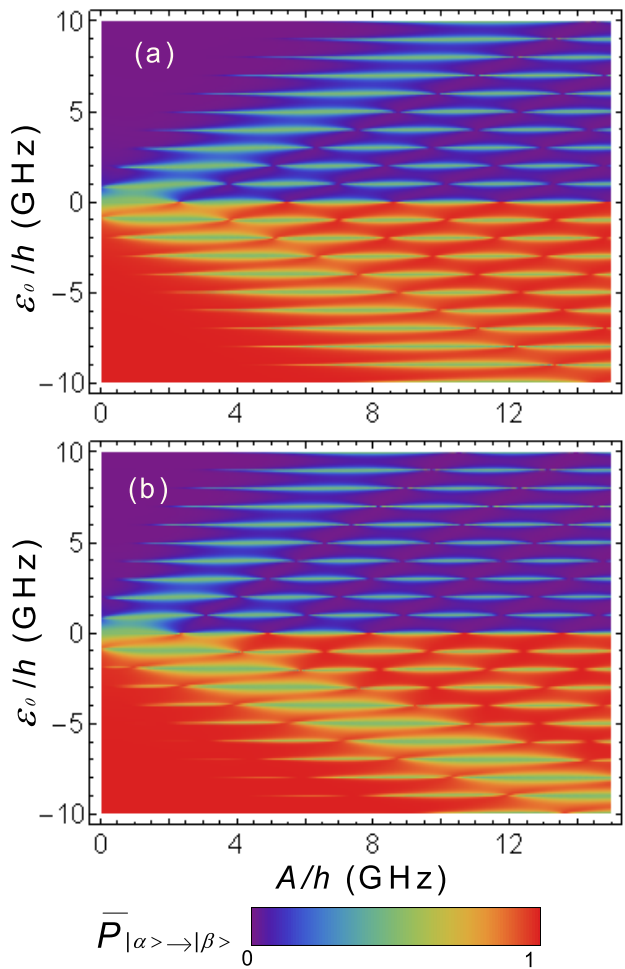}
\end{center}
\caption{\label{fig4} (color online) The population $\overline{P}_{ |\alpha\rangle\rightarrow |\beta\rangle}$ of the excited state $|\beta\rangle$ as function of the external field amplitude $A$ and the static bias $\varepsilon_{0}$ at $\gamma = 0$ (a) and $\gamma = 0.5$ (b). The qubit parameters used here are: $\Delta/h = 0.5$ GHz, $\omega/2\pi = 1$ GHz, and $\theta = \pi$.}
\end{figure}

Now, we will analyze in detail the above-mentioned features of the transition probabilities and how these depend on the driving field parameters. As in the preceding section, we investigate the population behavior of the qubit excited state $|\beta\rangle$ after changing the drive parameters, as done in amplitude spectroscopy \cite{Oliver,Berns,Sillanpaa,Berns1,Rudner}. Calculated according to Eq.~(\ref{15}), Fig.~\ref{fig4} shows the probability $\overline{P}_{|\alpha\rangle\rightarrow |\beta\rangle}$ for populating the state $|\beta\rangle$ versus the control parameter $\varepsilon_{0}$ and the amplitude $A$ of the external alternating field (at two values of the amplitude ratios $\gamma$ of the harmonic drive defined by Eq.~(\ref{2}) and for the relative phase $\theta = \pi $).

As mentioned above, these dependencies can be qualitatively understood in the context of the RWA. The oscillation frequency according to Eq.~(\ref{10}) is proportional to the sum of Bessel function products taken with different phases; therefore its minima and maxima are sensitive to the driving field parameters. The other peculiarity of this system is associated with the asymmetry over the off-set the static bias $\varepsilon_{0}$ which has been already discussed. Figure~\ref{fig4}(a) shows that for a monochromatic field ($\gamma = 0$), an interference pattern is symmetric with respect to $\varepsilon_{0}\rightarrow - \varepsilon_{0}$. Notice that such type of interference patterns have been obtained experimentally by using methods of amplitude spectroscopy \cite{Oliver,Berns,Sillanpaa,Berns1,Rudner}. Observed in Ref.~\onlinecite{Oliver,Berns,Sillanpaa,Berns1,Rudner} at $\gamma = 0$, the multi-photon qubit energy absorption is independent of the ``direction'' of the sweep over $\varepsilon_{0}$; while when $\gamma\neq0$ in Fig.~\ref{fig4}(b), the asymmetry in the location of the absorption peaks is clearly seen (see Ref.~\onlinecite{Bylander}). Figure~\ref{fig4}(b) also shows additional peaks caused by the form of the resonant condition $\varepsilon_{0}+(n+2m)\hbar\omega=0$ and by a set of $n$ and $m$ which are in close agreement with the absolute value, and which determine the dependence of the Bessel functions on the driving parameters. The interference pattern asymmetry allows, by changing the signal parameters, to control the Landau-Zener quantum-coherent tunneling and this could be important for controlling qubit states for large-amplitude drives.

%
\begin{figure}[b]
\begin{center}
      \includegraphics[width=6cm,height=3.5cm]{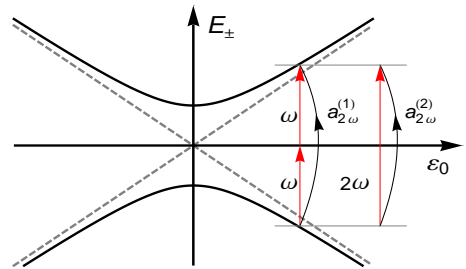}
\end{center}
\caption{\label{fig5} (color online) Energy diagram showing the eigenstates ($E_{+}$ and $E_{-}$) of a flux qubit as a function of the control parameter $\varepsilon_{0}$. Two components of the biharmonic drive may produce the transition pathways between the levels with amplitudes $a^{(2)}_{2\omega}$  and  $a^{(1)}_{2\omega}$, as shown in the figure.
}
\end{figure}
%
%

\section{THE RABI FREQUENCY DEPENDENCE ON RELATIVE PHASE AND AMPLITUDE}
%
%
%
%
%
\begin{figure*}
\begin{center}
      \includegraphics[width=13.0cm,height=12.5cm]{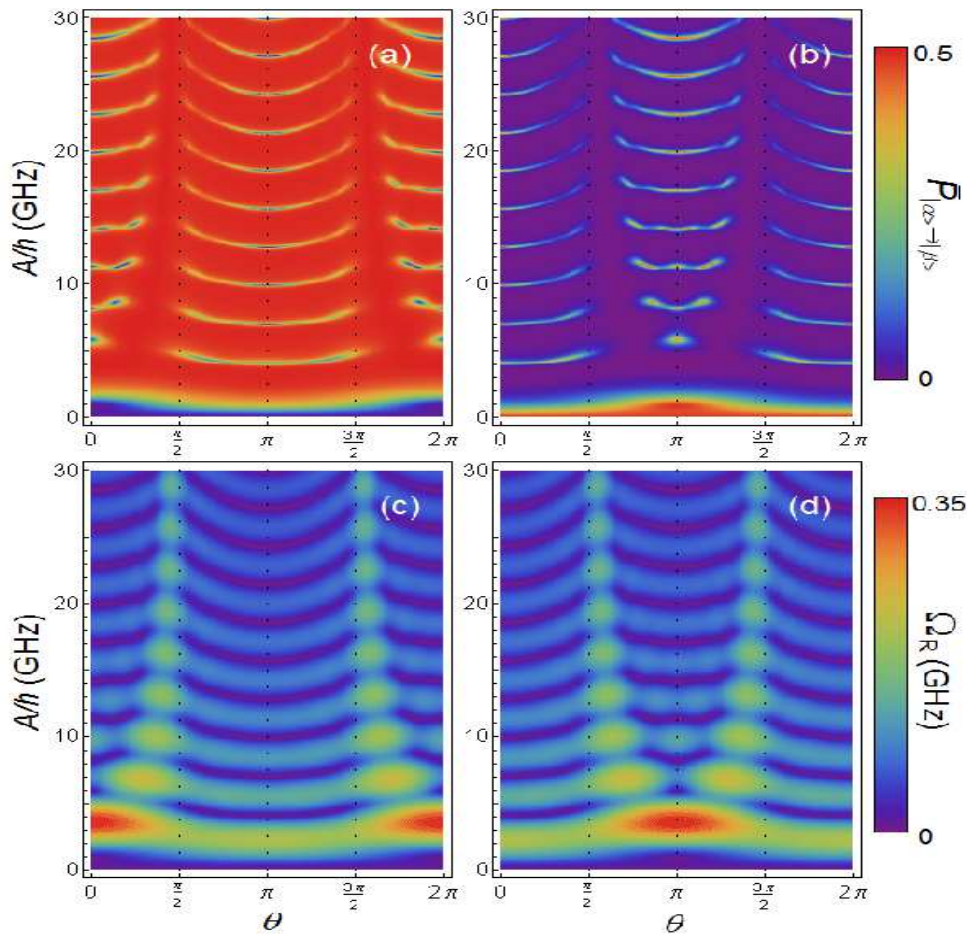}
\end{center}
\caption{\label{fig6} (color online) The probability $\overline{P}_{ |\alpha\rangle\rightarrow |\beta\rangle}$ of the excited state $|\beta\rangle$ depends on the relative phase $\theta$ and the amplitude $A$ of a biharmonic drive (a, b) and the Rabi frequency $\Omega_R$ in (c, d),  for $\varepsilon_{0}/h = 2$ GHz (a, c) and $\varepsilon_{0}/h = - 2$ GHz (b, d). The system parameters used here are: $\Delta/h = 0.5$ GHz, $\omega/2\pi = 1$ GHz, and $\gamma = 0.5$. On the right side of the figures the corresponding scales of the population probability $\overline{P}_{|\alpha\rangle\rightarrow |\beta\rangle}$ are given.}
\end{figure*}
%
%
We will concentrate here on the phase dependence of the level population of the excited qubit. In our case, the phase control arises by setting a relative phase difference  $\theta$ between the two components of the driving field.

Several features of the resonances, for the biharmonic driving Eq.(\ref{2}),  differs from the multiphoton resonance in monochromatic field.  Let us consider a biharmonic drive as a superposition of two weak drives arriving on a qubit. The nonlinear interaction of these biharmonic field with the qubit  produces  the harmonics. The perturbation approach, presented  in Sec. II,  shows that two harmonics may induce the transition pathways between qubits levels with the same frequencies.

%
%
%
\begin{figure*}
\begin{center}
       \includegraphics[width=16.0cm,height=12cm]{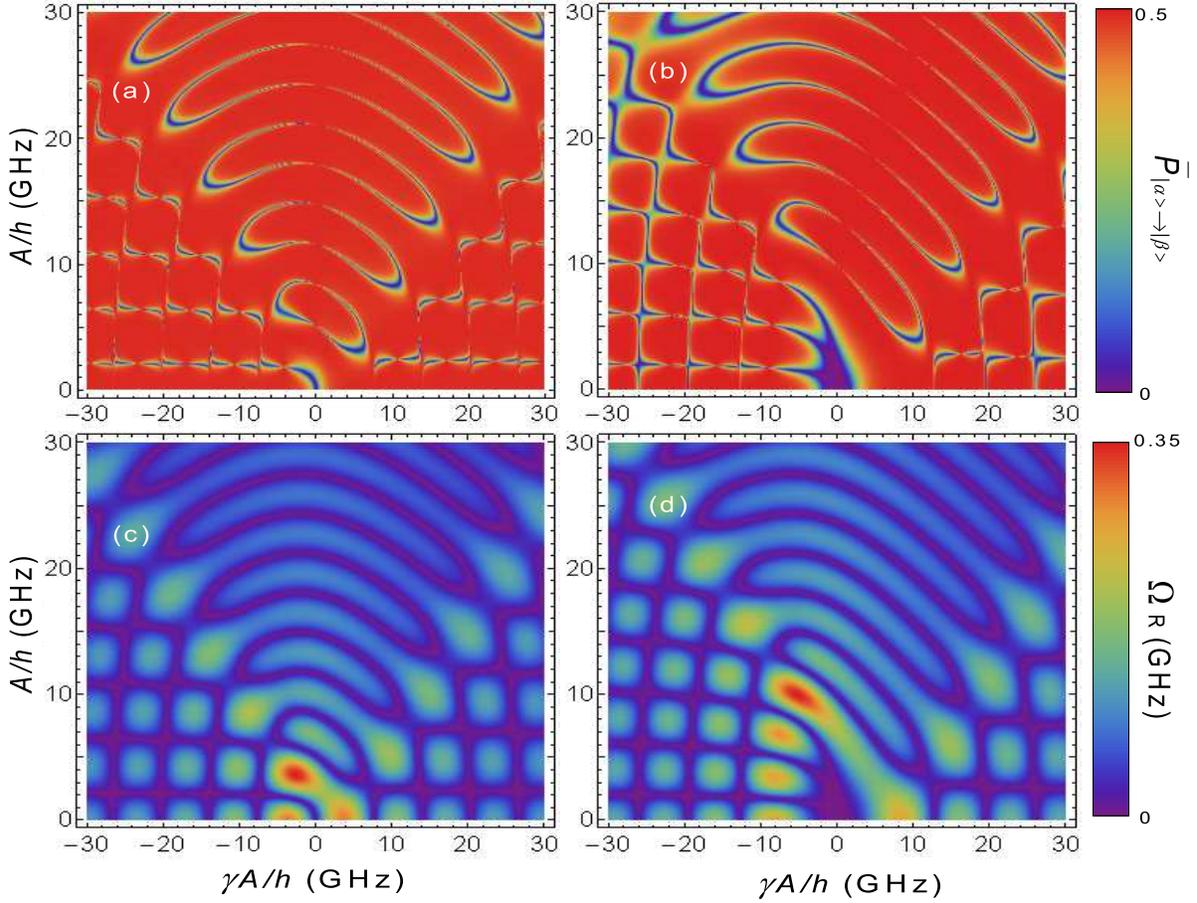}
\end{center}
\caption{\label{fig7} (color online) The probability $\overline{P}_{ |\alpha\rangle\rightarrow |\beta\rangle}$ of the excited state $|\beta\rangle$ versus the driving amplitudes $A$ and $\gamma A$ (a, b) and the Rabi frequency $\Omega_R$ in (c, d). Here, $\varepsilon_{0}/h = 2$ GHz in (a, c) and $\varepsilon_{0}/h = 6$ GHz in (b, d). The system parameters used here are: $\Delta/h = 0.5$ GHz, $\omega/2\pi = 1$ GHz, and $\theta = \pi$.}
\end{figure*}
%
For instance, one harmonic ($\sim\!A\cos(\omega t)$) with frequency $\omega$ can be transformed to a drive with frequency $2\omega$ ($\sim\!A^{2}\cos(2\omega t)$). This drive gives the transition between the qubit's levels with drive amplitude $a^{(2)}_{2\omega}$. At the same time, the harmonic $\sim\!\gamma A\cos(2\omega t + \theta)$ can cause a transition with the amplitude $a^{(1)}_{2\omega}$. This means that in this case it becomes possible to have a transition with probability $|a^{(2)}_{2\omega}+a^{(1)}_{2\omega}|^{2}$ and the interference population is caused by the nonlinear mixing of driving components on the qubit [see Fig.~\ref{fig5}]. Therefore, the mixing of two drives with different phases will be the result of  the phase dependence of the qubit population.

Figure~\ref{fig6} shows the probability to have a $|\alpha\rangle$ to $|\beta\rangle$ transition as a function of the relative phase difference, $\theta$, of the drives. Figure~\ref{fig6}(b) shows that when changing the sign of the controlling parameter $\varepsilon_{0}$, the probability $\overline{P}_{|\alpha\rangle\rightarrow|\beta\rangle} \approx 0.5$ is observed (the blue zones correspond to the appearance of a population plateau) and this is associated with the measured projection of the  current in a qubit.  The resonances of the Hamiltonian (\ref{9}) for a biharmonic drive are sensitive to the phase as illustrated in Fig.~\ref{fig6}(a). Figure~\ref{fig6} shows the locations of the maxima and minima, which are established by the transformation: $\varepsilon_{0}\rightarrow-\varepsilon_{0}$ and $\theta\rightarrow\theta+\pi s$, where $s$ is any integer. There are also special intervals of a relative phase difference (nearby $\theta=\pi/2$ and $\theta=3\pi/2$), when the Rabi frequencies are weakly-dependent on the field amplitude and the populations of  the excited state become constant. This population trapping effect can allow the dynamic control of the qubit. Indeed, for certain biharmonic field parameters it is possible to stabilize the population of a qubit in an excited state, and for  small changes of the signal amplitude the population remains stable.

To obtain additional information about how the shape of the biharmonic drive affects the qubit behavior, we computed the interference patterns of the excited state population [see Figs.~\ref{fig7}(a, b)] when changing the amplitudes $A$ and $\gamma A$, introduced in Eq.~(\ref{2}), respectively, with frequencies $\omega$ and $2\omega$. The blue zones in the red background refer to the capture of the population for the given parameters. The effect of the dynamical suppression of tunneling \cite{Grossmann} is seen in Figs.~\ref{fig7}(a, b); it occurs when the blue zone increases (i.e., the absence of excitation in the system) when changing the control parameter $\varepsilon_{0}$. The Rabi frequencies of the levels (colored) depend on  $A$ and $\gamma A$, and are shown to interpret the interference picture in Figs.~\ref{fig7}(c, d). It is possible to say that Figs.~\ref{fig7}(c, d) show the trajectories of the motion of the population zeroes which are shown in blue. In the RWA, as it is seen in Figs.~\ref{fig7}(c, d), the frequency of the Rabi generalized resonance qualitatively follows the behavior of the population zeroes in Figs.~\ref{fig7}(a, b).

When $\varepsilon_{0}=0$, a symmetric pattern of probabilities is formed along the axis $A\gamma = 0$, and when increasing the distance between the levels, the pattern deforms and a ``slope'' is observed. Note two significantly different zones of the resonance curves. First, a network of resonances in the right and left angles on the bottom zone of the squares in Figs.~\ref{fig7}(a, b) (for $|\gamma A| \gg A$). Second, the central zone has a divergent ``radial'' structure following the trajectories of the zeros of Bessel functions. The structure of network zones can be explained by the asymptotic behavior of the Bessel functions for large arguments ($\frac{A}{\hbar\omega}\gg\left|n^{2}-\frac{1}{4}\right |$ and $\frac{\gamma A}{2\hbar\omega}\gg\left|m^{2}-\frac{1}{4}\right |$) in the formula (\ref{10}) for the Rabi frequency:
%
%
\begin{widetext}
\[
J_{n}\left(\frac{A}{\hbar\omega}\right )J_{m}\left(\frac{\gamma A}{2\hbar\omega}\right )\approx\frac{2\hbar\omega}{\pi A}\sqrt{\frac{2}{\gamma}}\Bigl\{\cos\Bigl[\frac{A}{\hbar\omega}\left(1-\frac{\gamma}{2}\right)-\frac{\pi}{2}(n-m)\Bigr ]+
\sin\Bigl [\frac{A}{\hbar\omega}\left(1+\frac{\gamma}{2}\right)-\frac{\pi}{2}(n+m)\Bigr ]\Bigr\},
\]
\end{widetext}
which explains the formation of a periodic lattice. In the central zone, where $|\gamma A| \ll A$, in Eq.~(\ref{10}) a small amount of the components with $J_{m}\left (\frac{\gamma A}{2\hbar\omega}\right )$ is presented, so the zeroes of the Rabi frequency are basically determined by several Bessel functions.

Let us indicate one more  system symmetry which follows from the analysis of the Hamiltonian (\ref{1}): the Rabi frequency shows similar oscillations when $\varepsilon_{0}\rightarrow-\varepsilon_{0}$ and $\gamma\rightarrow-\gamma$, which corresponds to changing the sign in front of the harmonic with double frequency and points to the already-observed symmetry in the shift of the relative phase difference, $\theta \rightarrow \theta+\pi s$, where $s$ is any integer (see Fig.~\ref{fig6}).

The results presented in this section explore various  ways to control qubits by using the relative amplitude and the phase of a biharmonic signal.

\section{DEPHASING EFFECTS ON THE INTERFERENCE PATTERNS}

We finally briefly discuss the dephasing effects on the  qubit  interference patterns. Of course, in experimental conditions the interaction of a qubit with a reservoir (e.g., charge fluctuations on Josephson contacts, flux fluctuations through a superconducting circuit, and radiative damping) have a considerable effect on the qubit dynamics. These processes are typically described by considering the interaction of a qubit with a bosonic reservoir  \cite{Gardiner}. In this case, the equation for the density operator of the qubit $\rho$ in the Markov approximation takes the following form~\cite{Gardiner}
%
%
\begin{equation}
\frac{\partial \rho}{\partial t}=\frac{1}{i\hbar}\left [H,\rho\right ]+\frac{\Gamma}{2}\left (\sigma_{z}\rho \sigma_{z}-\rho \right ), \label{16}
\end{equation}
where the rate $\Gamma$ characterizes the phase damping and is determined by the reservoir parameters. The transverse relaxation (dephasing) usually dominates over the energy relaxation, which in this approximation can be neglected \cite{Oliver,Berns,Sillanpaa,Berns1,Rudner,Gardiner}.

%
%
%
%
\begin{figure}
\begin{center}
     \includegraphics[width=8cm,height=15cm]{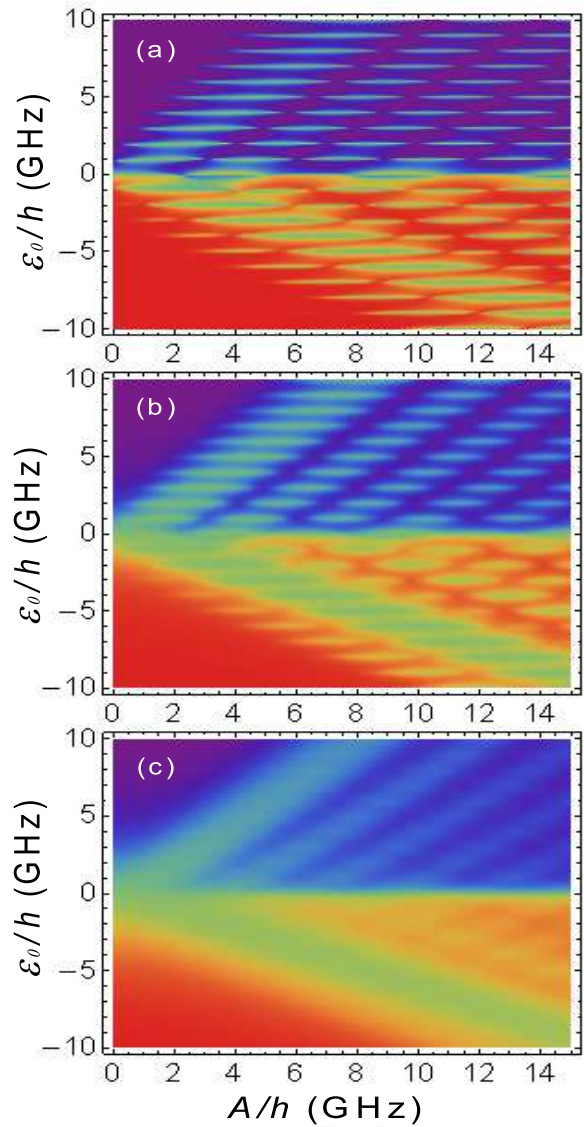}
\end{center}
\caption{\label{fig8} (color online) The probability of populating the excited state $|\beta\rangle$ after applying a biharmonic drive. Here  $\Delta/h = 0.5$ GHz, $\omega/2\pi = 1$ GHz, and the damping parameters are $\Gamma = 0.01$ GHz (a), $\Gamma = 0.09$ GHz (b), and $\Gamma = 0.36$ GHz (c). The scale of the transition probability is the same as in Fig.~\ref{fig4}. }
\end{figure}
%
%
According to Ref.~\onlinecite{Oliver}, dephasing produces broader and overlapping resonances already at $\Gamma\approx\omega/2\pi$, which also happens when the qubit is driven by a biharmonic drive (Fig.~\ref{fig8}). However, the asymmetry of the interference picture with respect to $\varepsilon_{0}$ and the population oscillation over the amplitude $A$ at a fixed $\varepsilon_{0}$ remains. Another difference is the slope of the interference fringes along the $A$ axis, which remains when $\Gamma\gg\omega/2\pi$. This can be used for the dynamic control of a qubit state fitting the phase difference between the two harmonics and their relative intensity.

\section{CONCLUSION}
The dynamic behavior of a qubit in a strong field depends significantly on the shape of the driving field. Let us briefly summarize a few results found here when a biharmonic field is used to drive a qubit. First, crossing the quasi-energy levels depends on the biharmonic drive parameters, causing a change of the multiphoton transition character according to the sign of the controlling parameter.
We have shown that the peaks of the resonances depend on the relative phase and amplitudes of the two harmonics driving the qubit.
Second, the interference pattern for the populations of a qubit in the excited state is sensitive to the driving field and noise parameters. These effects manifest the sensitivity of the level populations to the relative  phase. It is demonstrated that when the phases $\theta$ are multiples of $\pi/2$, the dynamical confinement of the populations are possible when changing the amplitudes of the drive. This effect can be used for the quantum control of the states of the qubit. The interference effects we obtained agree qualitatively with the results of experiments \cite{Bylander}.

Earlier we mentioned the analogy of forming Landau-Zener-St\"{u}eckelberg interference patterns of the qubit populations using a Mach-Zehnder interferometer \cite{Born,Shevchenko1,Petta} (see Ref.~\onlinecite{Oliver} for example). According to this analogy, the qubit evolves differently in the upper and lower levels and (Landau-Zener) transitions occur when the levels approach each other. The interference of two states propagating along two levels causes the formation of the interference pattern. The Landau-Zener tunneling can be seen as similar to the passage of light through semitransparent mirrors. In the case of a biharmonic drive, the interference pattern depends on the form of the driving field. Following this analogy, the light beams meet two types of mirrors and their permeability (tunneling probability) through the regions of adiabatic level-crossing (and consequently the interference pattern of the excited-state population) become sensitive to the form of the driving field.

\acknowledgments
We are very grateful to W.D. Oliver  for a careful reading of the manuscript and  helpful  remarks. FN acknowledges partial support from the ARO, Grant-in-Aid
for Scientific Research (S), MEXT Kakenhi on Quantum
Cybernetics, the JSPS via its FIRST program and the JSPS-RFBR Grant No. 12-02-92100.
This work was funded in part by the Russian Ministry of Education and Science through the programs No. 07.514.11.4147; No. 14.132.21.1399, and the RFBR Grant No. 12-07-00546; No. 12-07-31144. M.V.D. was financially supported by the ``Dinastia'' fund.

\end{document}